\documentclass[conference,twoside,times,]{IEEEtran}
\usepackage{cite,graphicx,amsmath,amsfonts,amssymb}

\newtheorem{theorem}{Theorem}

\newtheorem{definition}{Definition}
\newtheorem{corollary}{Corollary}

\renewcommand{\QED}{\QEDopen}

\begin{document}

\title{A Class of LDPC Erasure Distributions with Closed-Form Threshold Expression}

\author{\authorblockN{Enrico Paolini and Marco Chiani}
\authorblockA{DEIS, WiLAB\\
University of Bologna\\
via Venezia 52, 47023 Cesena (FC), Italy\\
Email: \{epaolini, mchiani\}@deis.unibo.it}}
\date{\today}
\pagestyle{empty} \thispagestyle{empty} \setcounter{page}{1}
\maketitle
%
\begin{abstract}
In this paper, a family of low-density parity-check (LDPC) degree
distributions, whose decoding threshold on the binary erasure
channel (BEC) admits a simple closed form, is presented. These
degree distributions are a subset of the check regular
distributions (i.e. all the check nodes have the same degree), and
are referred to as $p$-positive distributions. It is given proof
that the threshold for a $p$-positive distribution is simply
expressed by $[\lambda'(0)\rho'(1)]^{-1}$. Besides this closed
form threshold expression, the $p$-positive distributions exhibit
three additional properties. First, for given code rate, check
degree and maximum variable degree, they are in some cases
characterized by a threshold which is extremely close to that of
the best known check regular distributions, under the same set of
constraints. Second, the threshold optimization problem within the
$p$-positive class can be solved in some cases with analytic
methods, without using any numerical optimization tool. Third,
these distributions can achieve the BEC capacity. The last
property is shown by proving that the well-known binomial degree
distributions belong to the $p$-positive family.
\end{abstract} {\pagestyle{plain} \pagenumbering{arabic}}
%
%
\section{Introduction}\label{introduction}
The unavailability of a closed form expression for the decoding
threshold of low-density parity-check (LDPC) codes still
represents an open problem. So far, no general closed form
threshold expression is known for any transmission channel. Some
results in this sense have been developed for the binary erasure
channel (BEC), for which some analytic threshold expressions have
been proposed. In \cite{bazzi04:exact}, an analytic threshold
expression has been presented for regular LDPC ensembles. A more
general analytic expression, valid for check regular ensembles,
has been proposed in \cite{paolini05:threshold}. To the best of
the authors' knowledge, the most general analytic expression
currently available is that one developed in \cite{hehn05:exact},
which can be applied to fully irregular ensembles. For regular or
check regular ensembles, the formula proposed in
\cite{hehn05:exact} coincides with that one from
\cite{paolini05:threshold}. The problem with these expressions is
that none of them can be really considered a \emph{closed form}
one (except for the case of regular LDPC ensembles with degree-2
variable nodes presented in \cite{bazzi04:exact}). In fact, in all
these formulas, the threshold is a function of some parameter
which in general does not admit a closed form expression. This
parameter, which depends on the degree distribution, can be a root
of a real polynomial, a fixed point of a real function, or the
abscissa of the tangent point between two EXIT curves
\cite{ten-brink04:design}.

An exception is represented by the degree distributions for which
the first occurrence of a tangency point between the EXIT curves
in the EXIT chart appears for a value of the {\em a priori} mutual
information equal to 1 (this is sometimes referred to as
derivative matching condition). Denoting the decoding threshold by
$\delta^*$, these distributions achieve with equality the
stability condition $\delta^* \leq [\lambda'(0)\rho'(1)]^{-1}$
\cite{richardson01:design} (which holds for any distribution).
This paper presents a family of check regular LDPC distributions,
called $p$-positive distributions, which are a subset of the
distributions achieving the derivative matching condition. The
starting point for obtaining this family is the afore mentioned
analytic formula developed in \cite{paolini05:threshold}.

The name ``$p$-positive" is chosen because these distributions are
defined as those ones for which a certain polynomial $p\,(\cdot)$,
whose coefficients depend on the edge-oriented variable
distribution $\lambda(x)$ and
on the degree $d_c$ of the check nodes, is 
non-negative between 0 and 1. The theory of polynomials can be
applied to obtain conditions for a check regular distribution to
be $p$-positive. A useful theorem in this sense is the
Fourier-Budan theorem \cite[p. 27]{barbeau03:polynomials}, which
states an upper bound to the number of real roots in a given
interval, for a polynomial with real coefficients. The application
of the Fourier-Budan theorem to the polynomial $p\,(\cdot)$ leads
to a set of inequalities involving the coefficients $\lambda_i$'s
and $d_c$, which represent a necessary condition for a
distribution to be $p$-positive.

Besides 
the closed form threshold expression, the $p$-positive
distributions are shown to exhibit 
some additional good properties. First, the threshold of the
optimal $p$-positive distribution under a set of constraints,
represented by the given code rate, check degree and maximum
variable degree, is in some cases extremely close to that of the
best known check regular degree distributions. Second, within the
$p$-positive class, it is in some cases possible to optimize the
degree
distribution with analytic methods only, i.e. without using 
numerical optimization tools. In fact, despite being only
necessary, the condition obtained from the Fourier-Budan theorem
is shown to be sufficient, in some cases, to find the $p$-positive
distribution with optimum threshold under the given set of
constraints. Third, it is proved that capacity achieving sequences
exist within the $p$-positive class. More specifically, it is
recognized that any binomial degree distribution
\cite{shokrollahi99:new} is $p$-positive.

The paper is organized as follows. Section
\ref{section:right-regular-threshold} recalls the threshold
expression for check regular ensembles from
\cite{paolini05:threshold}. Section
\ref{section:p-positive} exploits this formula for introducing the new class of 
distributions, and develops a necessary condition for a check
regular distribution to belong to this class. Section
\ref{section:optimization} is devoted to the threshold
investigation of the $p$-positive distributions, with both
differential evolution \cite{shokrollahi00:design} and analytic
methods. In Section \ref{sec:binomial}, the existence of capacity
achieving sequences within the $p$-positive family is proved.
Section \ref{section:conclusion} concludes the paper.

%
\section{Threshold for Check Regular Distributions}\label{section:right-regular-threshold}
%
Let $\mathcal{C}^{n}(\lambda,\rho)$ be the ensemble of all the
length-$n$ LDPC codes with edge-oriented degree distribution
$(\lambda,\rho)$ \cite{richardson01:design}. Let $\delta$ be the
BEC erasure probability. Then, the (bit oriented) threshold
$\delta^*$ for the ensemble $\mathcal{C}^{n}(\lambda,\rho)$ on the
BEC is defined as the maximum $\delta$ for which the residual bit
erasure probability can be made arbitrarily small by increasing
the number of decoding iterations, in the limit where the codeword
length $n$ tends to infinity.

Density evolution for the BEC \cite{richardson01:design}, can be
expressed as follows. If $x_\ell$ is the probability that a
message from a variable node to a check node during the $\ell$-th
decoding iteration is an erasure message and the bipartite graph
is cycle-free (infinite codeword length), then $x_{\ell+1}= \delta
\, {\lambda}(1- \rho (1-x_{\ell} ))$. Hence, the threshold
$\delta^*$ is the maximal $\delta$ for which $x_\ell \rightarrow
0$ in the limit where $\ell \rightarrow \infty$. The first
appearance of a non-zero fixed point for $x_\ell$ is a necessary
and sufficient condition for the corresponding $\delta$ to be the
threshold.

From this observation, $\delta^*$ can be equivalently expressed as
the maximal $\delta$ for which $ \delta \, \lambda(1 - \rho(1 -
x)) < x$ $\forall \: x \, \in (0,\delta] $
\cite{luby97:practical}, or the maximal $\delta$ for which:
\begin{align}\label{eq:de-inequality-2}
\rho(1 - \delta \, \lambda(x)) > 1 - x \qquad \forall \: x \, \in
(0,1] \, .
\end{align}
For check regular codes, with variable distribution
$\lambda(x)=\sum_{i \geq 2} \lambda_i x^{i-1}$ and check
distribution $\rho(x)=x^{d_c-1}$ ($d_c \geq 3$), the inequality
\eqref{eq:de-inequality-2} becomes:
\begin{align}\label{eq:de-right-regular}
(1-\delta\lambda(x))^{d_c-1}>1-x \qquad \forall \: x \, \in (0,1].
\end{align}

Let us define $f(x,\delta) = (1-\delta\lambda(x))^{d_c-1}$ and
suppose at first $\lambda_2 = 0$. For any given $x \in (0,1]$,
$f(x,\delta)$ is continuous and monotonically decreasing with
respect to $\delta$, varying from $f(x,0)=1$ to
$f(x,1)=(1-\lambda(x))^{d_c-1} > 0$. Moreover, it is everywhere
derivable for $x$ in $(0,1)$, and the following relationships
hold:
\begin{align*}
\partial f / \partial x \, (0,\delta) & = \, 0 \quad \forall \,
\delta \in (0,1)\\
f(0,\delta) & = \, 1 \quad \forall \, \delta \in (0,1)\\
f(1,\delta) & = \, (1-\delta)^{d_c-1} > 0\\
f(x,0) & = \, 1 \quad \forall \, x \in (0,1).
\end{align*}
It follows from these properties that the maximal value of
$\delta$ for which \eqref{eq:de-right-regular} holds is the
minimum value of $\delta$ for which the graph of $f(x,\delta)$
(considered as a function of $x$ with $\delta$ playing the role of
parameter) is tangent to the graph of $g(x)=1 - x$, for some
$x=\gamma$. The tangency condition in $x = \gamma$ is:
\begin{equation}\label{eq:tangent-right-regular}
\left\{
\begin{array}{l}
(1-\delta\lambda(\gamma))^{d_c-1} = 1-\gamma\\
\delta(d_c-1)\lambda'(\gamma)(1-\delta\lambda(\gamma))^{d_c-2} =
1.
\end{array} \right.
\end{equation}

The relationships \eqref{eq:tangent-right-regular} are not able to
unambiguously determine the threshold. In fact,
\eqref{eq:tangent-right-regular} can admit several discrete
solutions $(\delta,\gamma)$, with $\gamma \in (0,1)$ and $\delta
\in (0,1)$. From a geometrical point of view, several discrete
values of $\delta \in (0,1)$ can exist for which the graph of
$f(x,\delta)$ (for fixed $\delta$) is tangent to the graph of
$g(x)=1-x$ in some $x = \gamma$. The threshold $\delta^*$ for the
check regular distribution is the minimum among these discrete
values of $\delta$.

The second equation of \eqref{eq:tangent-right-regular} can be
written as
$$
(1-\delta\lambda(\gamma))^{d_c-2}=[\delta (d_c-1)
\lambda'(\gamma)]^{-1} \,  ,
$$
that can be substituted in the first equation in order to obtain
\begin{align}\label{eq:delta-expression-right-regular}
\delta = [\lambda(\gamma)+(d_c-1)(1-\gamma)\lambda'(\gamma)]^{-1}
\,  .
\end{align}
In the following, $h(\cdot)$ will denote the following key
function:
\begin{align}\label{eq:h-definition}
h(x) & = [\lambda(x)+(d_c-1)(1-x)\lambda'(x)]^{-1}.
\end{align}
By substituting \eqref{eq:delta-expression-right-regular} into the
first equation of \eqref{eq:tangent-right-regular}, we obtain
$$
\frac{\lambda(\gamma)+(d_c-1)(1-\gamma)\lambda'(\gamma)}{(d_c-1)(1-\gamma)\lambda'(\gamma)}
= (1-\gamma)^{-\frac{1}{d_c-1}},
$$
and developing this expression leads to:
\begin{align}\label{q:gamma-equation-right-regular}
\gamma = 1 -
\Big(\frac{\lambda(\gamma)}{(d_c-1)\lambda'(\gamma)}+1-\gamma\Big)^{\frac{d_c-1}{d_c-2}}.
\end{align}

Summarizing, for a given check regular ensemble with $\lambda_2 =
0$, the threshold on the BEC is equal to the minimum $h(\gamma)$
under the constraint that $\gamma$ is one of the (usually) several
fixed points in $(0,1)$ of
\begin{align}\label{eq:function-phi-definition}
\phi(x)=1 -
\Big(\frac{\lambda(x)}{(d_c-1)\lambda'(x)}+1-x\Big)^{\frac{d_c-1}{d_c-2}}
\, .
\end{align}
Note that $\gamma=0$ is always a fixed point for $\phi(\cdot)$.
However, for $\lambda_2=0$ the threshold is never achieved in
correspondence of $\gamma=0$, because in this case $\lim_{x
\rightarrow 0+} h(x) = +\infty$.

The only differences when removing the hypothesis $\lambda_2 = 0$
are that $\partial f / \partial x(0,\delta) < 0 \,\,\, \forall \,
\delta \in (0,1)$, and that $h(0)$ is finite and equal to
$[\lambda_2 (d_c-1)]^{-1} = [\lambda'(0)\rho'(1)]^{-1}$. Hence,
the threshold could be achieved in correspondence of the fixed
point $\gamma = 0$.
Thus, the threshold for a check regular ensemble with $\lambda_2
> 0$ is equal to the minimum $h(\gamma)$, with $\gamma$ fixed
point in $[0,1)$ of $\phi(\cdot)$.

%
\medskip
In the special case of regular LDPC codes, with variable
distribution $\lambda(x)=x^{d_v-1}$, $d_v \geq 3$, and check
distribution $\rho(x)=x^{d_c-1}$, $d_c \geq 3$, the threshold is
given by
\begin{equation}\label{threshold-expression-regular}
\delta^* = \frac{[(d_c-1)(d_v-1)]^{-1}}{\gamma^{d_v-2} - c \,
\gamma^{d_v-1}} \, ,
\end{equation}
where $\gamma$ is the unique fixed point in $(0,1)$ of
\begin{equation}\label{psi}
\psi(x) = 1 - (1 - c \, x)^{\frac{d_c-1}{d_c-2}} \, .
\end{equation}
with $c = [(d_c-1)(d_v-1)-1]/[(d_c-1)(d_v-1)] < 1$.
%
%
\section{The New Class of Degree Distributions}\label{section:p-positive}
In this section, the class of $p$-positive degree distributions is
introduced.

\medskip
\begin{definition}
A degree distribution with code rate $R$ and threshold $\delta^*
 =(1 - \epsilon)(1 - R)$ is called \emph{$(1-\epsilon)$ capacity
achieving of rate $R$}.
\end{definition}

\medskip
Consider a check regular distribution for which $h(x) \geq h(0)$
$\forall x \in (0,1]$, where $h(\cdot)$ is defined in
\eqref{eq:h-definition}. This can be equivalently written as
$[h(x)]^{-1} \leq [h(0)]^{-1}$, i.e.
$$\sum_{i=2}^{L} \lambda_i x^{i-1} + (d_c-1)(1-x)\sum_{i=2}^{L}(i-1) \lambda_i x^{i-2} \leq \lambda_2
(d_c-1),$$
where $L$ denotes the maximum variable degree. After some
algebraic manipulation, the previous inequality can be put in the
form $p(x) \geq 0$, where $p\,(\cdot)$ is the real polynomial
\begin{align}\label{eq:p(x)-definition}
p\,(x) = \omega_L \, \lambda_L \, x^{L-2} + \sum_{i=2}^{L-1}
[\omega_i \lambda_i - (\omega_{i+1}+1) \, \lambda_{i+1}] \,
x^{i-2},
\end{align}
where $\omega_i = (d_c-1)(i-1)-1$. The polynomial $p\,(\cdot)$
will be expressed in the form $p\,(x) = \sum_{i=0}^{L-2}p_i \,
x^i$. The condition $h(x) \geq h(0)$ for all $x \in (0,1]$ is
equivalent to the condition that the real polynomial $p\,(\cdot)$
is positive or null in $(0,1]$.
The class of degree distribution pairs studied in this paper is
introduced next.

\medskip
\begin{definition}
A \emph{$p$-positive distribution} is any check regular degree
distribution with $\lambda_2 > 0$, such that $p(x) \geq 0$
$\forall$ $x \in (0,1]$.
\end{definition}

\medskip
The following theorem individuates a simple closed form for the
threshold of the $p$-positive distributions, and states the
necessary and sufficient condition for any $p$-positive
distribution to be $(1-\epsilon)$ capacity achieving of rate $R$.
\medskip
%
%
\begin{theorem}\label{theo:lambda-2}
The threshold of any $p$-positive degree distribution is equal to
$[\lambda'(0) \rho'(1)]^{-1}$. The degree distribution is
$(1-\epsilon)$ capacity achieving of rate $R$ if and only if
\begin{equation}\label{eq:lambda2-theorem}
\lambda_2 = [(1 - \epsilon)(1 - R)(d_c - 1)]^{-1}.
\end{equation}
\end{theorem}
\medskip
\begin{proof}
For a $p$-positive distribution, $h(x) \geq h(0)$ for all $x \in
(0,1]$. From Section \ref{section:right-regular-threshold} it is
known that the threshold for a check regular ensemble with
$\lambda_2 > 0$ is the minimum among discrete values $h(\gamma)$,
with $\gamma$ fixed point in $[0,1)$ of $\phi(\cdot)$ defined in
\eqref{eq:function-phi-definition}. Moreover, $\gamma = 0$ is
always a fixed point of $\phi(\cdot)$ under the condition
$\lambda_2 > 0$. If $h(x) \geq h(0)$ for all $x \in (0,1]$, then
the minimum is always achieved in correspondence of the fixed
point $\gamma = 0$, independently of the number and the positions
of all the other fixed points of $\phi(\cdot)$. Hence, $\delta^* =
h(0) = [\lambda'(0)\rho'(1)]^{-1} = [\lambda_2 (d_c-1)]^{-1}$. If
and only if equality \eqref{eq:lambda2-theorem} holds, it is
$\delta^* = (1 - \epsilon)(1 - R)$.
\end{proof}
%
%

\medskip
For a given $\epsilon > 0$, the search for $p$-positive
$(1-\epsilon)$ capacity achieving of rate $R$ distributions, with
check degree $d_c$ and maximum variable degree $L$, can be
performed by letting $\lambda_2$ be expressed by
\eqref{eq:lambda2-theorem}, and looking for $\lambda_i$, $i = 3,
\dots,L $, such that $p\,(x) \geq 0$ for all $x \in (0,1]$. In
general, for some $R$, $d_c$ and $L$, this problem admits
solutions for $\epsilon \geq \epsilon_{\textrm{opt}}>0$, with
$\epsilon_{\textrm{opt}}$ depending on $R$, $d_c$ and $L$. The
value $\epsilon_{\textrm{opt}}$ is associated to the optimal
$p$-positive distribution, under the imposed set of constraints
(i.e. code rate, check nodes degree and active variable degrees).

It is readily shown that $p\,(1)=(d_c-1)\lambda_2-1$, which is
positive (from \eqref{eq:lambda2-theorem}). Then, the condition
that $p\,(x) \geq 0$ for $x$ between 0 and 1, is equivalent to the
condition that $p\,(\cdot)$ has only roots with even multiplicity
between 0 and 1. Several well known properties of the real roots
of polynomials can be then exploited, in order to develop
conditions for a check regular distribution to be $p$-positive. In
particular the following theorem, known as the Fourier-Budan
theorem, permits to obtain a simple necessary (but not sufficient)
condition. It provides an upper bound to the number of zeroes of a
real polynomial between two values $a$ and $b$, with $a<b$.

\medskip
%
%
\begin{theorem}[Fourier-Budan Theorem \emph{\cite[p. 27]{barbeau03:polynomials}}$\,$]\label{theo:FB}
Let $p\,(\cdot)$ be a real polynomial of degree $q$, and let $a$
and $b$ be two real values such that $a<b$, and $p\,(a) \cdot
p\,(b) \neq 0$. Then, the number of real roots of $p\,(\cdot)$
between $a$ and $b$ (each one counted with its multiplicity) is
\emph{not greater} than $A-B$, where $A$ and $B$ are,
respectively, the number of sign changes in sequences:
\begin{align*}
p\,(a), \, p'(a), \, p''(a), \, \dots, \, p^{(q)}(a)\\
p\,(b), \, p'(b), \, p''(b), \, \dots, \, p^{(q)}(b).
\end{align*}
Moreover, if the number of real roots of $p\,(\cdot)$ between $a$
and $b$ is smaller than $A-B$, then it differs from $A-B$ by an
even number.
\end{theorem}
%
%

\medskip
This theorem can be directly applied to the polynomial
$p\,(\cdot)$ defined by \eqref{eq:p(x)-definition}, where $a=0$
and $b=1$, leading to the following corollary.

\medskip
\begin{corollary}\label{coroll:FB-corollary}
The number of real roots between 0 and 1 of $p\,(x) =
\sum_{i=0}^{L-2}p_i \, x^i$ defined by \eqref{eq:p(x)-definition}
is not greater than the number $A$ of sign changes in the sequence
$p\,_0, \, p\,_1, \dots, \, p\,_{L-2}$. If the number of roots is
smaller than $A$, then it differs from $A$ by an \emph{even}
number.
\end{corollary}
\begin{proof}
The $i$-th derivative of $p\,(\cdot)$ in 0 and 1 is
\begin{align*}
p^{(i)}(0) & = i! \, p_i \notag \\
p^{(i)}(1) & = \sum_{j=i}^{L-2} \frac{j!}{(j-i)!} \, p_j.
\end{align*}
From the structure of the coefficients of $p\,(\cdot)$ it is not
difficult to recognize that all the values in the sequence
$p\,(1), \, p'(1), \, \dots, \, p^{(L-2)}(1)$ are positive for any
$L$ and $d_c \geq 3$. Hence, $B=0$ and the number of roots of
$p\,(\cdot)$ between 0 and $1$ is not greater than the number $A$
of sign changes in the sequence $p\,_0, \, 1! p\,_1, \dots,
(L-2)!\,p\,_{L-2}$. This is equal to the number of sign changes in
$p\,_0, \, p\,_1, \dots, \, p\,_{L-2}$. By directly applying
Theorem \ref{theo:FB}, we obtain the statement. We also observe
that $p\,_{L-2} = [(L-1)d_c - L] \, \lambda_{L}$ is always
positive.
\end{proof}

\medskip
Recall that, for the $p$-positive distributions, $p(\cdot)$ has
only roots, between 0 and 1, with even multiplicity. Then, the
following necessary condition for a check regular distribution to
be a $p$-positive distribution is obtained from Corollary
\ref{coroll:FB-corollary}.

\medskip
\begin{corollary}\label{coroll:necessary}
Necessary condition for a check regular distribution to be a
$p$-positive distribution is that the number of sign changes in
the sequence $p\,_0, \, p\,_1, \dots, \, p\,_{L-2}$ is even.
\end{corollary}
\medskip
In the next section, some examples of threshold optimizations for
$p$-positive distributions are provided. These results reveal
that, for some values of the code rate $R$, check degree $d_c$ and
maximum variable degree $L$, the $p$-positive distributions are
characterized by very good thresholds. In some cases, the
threshold is very close to that one of the best known
distributions, under the same set of constraints.

\section{Optimization of $p$-positive Degree
Distributions}\label{section:optimization}
\subsection{Optimization with Constrained Differential
Evolution}\label{subsec:DE-optimization} The threshold
optimization problem for the $p$-positive distributions can be in
principle performed numerically, using a constrained version of
the differential evolution (DE) algorithm. Specifically, a number
$N_D+1$ of values $x_i=(1/N_D) \cdot i$, for $i=0,\dots,N_D$, are
first chosen, for sufficiently large $N_D$, and then the DE
optimization tool is run for given code rate, given check degree
and given maximum variable degree, and with the additional set of
constraints $p\,(x_i) \geq 0$ for all $i$.

We performed this constrained DE optimization for $R=1/2$, $L=20$
and $d_c = \{6,7,8\}$. Under the same set of constraints, we also
performed the DE optimization, without imposing the $p$-positive
bound. The results of this search are shown in Table
\ref{tab:distributions-numerical} for $d_c=6$ and $d_c=7$. In both
cases, the best $p$-positive distribution exhibits a threshold
that is only slightly worse than that one of the best check
regular distribution obtained removing the $p$-positive
constraint, with the advantage represented by the closed form
threshold expression.

This conclusion is no longer valid for $d_c=8$. In fact, for this
check node degree, the best found $p$-positive threshold was
$\delta^*=0.469592$, while the best found check regular threshold
was $\delta^*=0.491988$. This means that the $p$-positive
constraint is not ``compatible" with all the possible sets of
bounds on the code rate, check degree and maximum variable degree.
When this compatibility holds, the $p$-positive distributions
exhibit very good thresholds. In the other cases, they exhibit a
threshold loss with respect to the best known distributions.

\subsection{Optimization Based on the Fourier-Budan Theorem}
In Section \ref{subsec:DE-optimization}, the $p$-positive
distribution optimization has been performed according to a
constrained version of the DE algorithm. Some examples of
$p$-positive threshold optimization, based on a totally different
technique, are presented next. More specifically, it is shown that
the necessary condition, developed in the previous section
(Corollary \ref{coroll:necessary}), can be exploited in order to
perform the optimization process without using numerical tools.
This is possible for particular structures of the variable nodes
degree distribution.

As a case study, let us consider check regular distributions with
variable distribution in the form $\lambda(x)=\lambda_2\,x +
\lambda_3\,x^2+\lambda_K\,x^{K-1}+\lambda_L\,x^{L-1}$ with
$L\geq7$ and $4<K<L-1$. For any choice of $R$, $d_c$, $K$ and $L$,
$\epsilon_{\textrm{opt}}$ will denote the minimum $\epsilon$ such
that a corresponding $p$-positive $(1-\epsilon)$ capacity
achieving distribution of rate $R$ exists, for the given
parameters.

For such $\lambda(\cdot)$, the coefficients of $p\,(\cdot)$ are:
\begin{align*}
\, & p\,_0 = (d_c-2)\lambda_2-(2d_c-2)\lambda_3\\
\, & p\,_1 = (2d_c-3)\lambda_3\\
\, & p\,_{K-3} = -[(K-1)d_c-(K-1)]\lambda_K\\
\, & p\,_{K-2} = [(K-1)d_c-K]\lambda_K\\
\, & p\,_{L-3} = -[(L-1)d_c-(L-1)]\lambda_L\\
\, & p\,_{L-2} = [(L-1)d_c-L]\lambda_L.\\
\end{align*}
According to Corollary \ref{coroll:necessary}, necessary (but not
sufficient) condition for the degree distribution to be
$p$-positive is that the number $A$ of sign changes in the
sequence $p\,_0,p\,_1,p\,_{K-3},p\,_{K-2},p\,_{L-3},p\,_{L-2}$ is
even. In the specific case under analysis, the following
inequalities are always true:
$p_{L-2}>0,\,p_{L-3}<0,\,p_{K-2}>0,\,p_{K-3}<0,\,p_1>0$. Then, if
and only if $p\,_0>0$, the condition that $A$ is even is satisfied
(specifically, it results $A=4$).

\begin{table}[!t]
\caption{Optimal $p$-positive and not $p$-positive distributions
and thresholds for $R=1/2$, $L=20$ and
$d_c=\{6,7\}$.}\label{tab:distributions-numerical}
\begin{center}
\begin{tabular}{|c|c|c|c|c|}
\hline \multicolumn{1}{|c|}{$\,$} & \multicolumn{2}{|c|}{$d_c=6$} & \multicolumn{2}{|c|}{$d_c=7$}\\
\hline\hline $\,$ & $p$-pos. & not $p$-pos. & $p$-pos. & not $p$-pos.\\
\hline
$\lambda_2$ & 0.415884 & 0.415273 & 0.339162 & 0.338843\\
$\lambda_3$ & 0.165968 & 0.160268 & 0.138401 & 0.140058\\
$\lambda_4$ & 0.095028 & 0.142202 & 0.104711 & 0.104198\\
$\lambda_5$ & 0.106071 &          & 0.033138 &         \\
$\lambda_6$ &          & 0.034597 &          & 0.087264\\
$\lambda_7$ &          &          & 0.166166 & 0.104669\\
$\lambda_8$ & 0.070638 & 0.247661 &          &         \\
$\lambda_9$ & 0.146412 &          &          &         \\
$\lambda_{14}$ &       &          & 0.104300 &         \\
$\lambda_{16}$ &       &          &          & 0.224968\\
$\lambda_{19}$ &       &          & 0.114122 &         \\
\hline \hline
\multicolumn{5}{|c|}{\emph{Thresholds}}\\
\hline
$\delta^*$ & 0.480904 & 0.481524 & 0.491407 & 0.491740\\
\hline
\end{tabular}
\end{center}
\end{table}

\begin{figure}[t]
\begin{center}
\includegraphics[width=8 cm, angle=0]{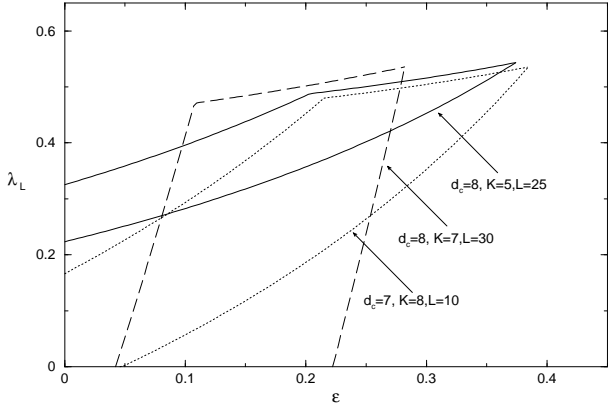}
\end{center}
\caption{Permitted region for some values of $(d_c,K,L)$.}
\label{fig:permitted-region}
\end{figure}

Suppose to fix the code rate $R$, the check nodes degree $d_c$ and
the active variable degrees ($K$ and $L$). From the constraints
$\sum_{i=2}^{L}\lambda_i=1$ and $R=1-d_c/(\sum_{i=2}^{L}
\lambda_i/i)$, and from \eqref{eq:lambda2-theorem}, expressions of
$\lambda_3$ and $\lambda_K$ as functions of $\epsilon$ and
$\lambda_L$, namely $\lambda_3 = \lambda_3(\epsilon,\lambda_L)$
and $\lambda_K = \lambda_K(\epsilon,\lambda_L)$, can be found.
Then, the necessary condition for the distribution to be
$p$-positive is given by the following set of inequalities:
\begin{align*}
\,& 0<\lambda_2(\epsilon)<1 \qquad
0<\lambda_3(\epsilon,\lambda_L)<1\\
\,& 0<\lambda_K(\epsilon,\lambda_L)<1 \qquad 0<\lambda_L<1\\
\,& p_0(\epsilon,\lambda_L)>0.
\end{align*}
The solution of this set of inequalities identifies a region on
the plane $\epsilon - \lambda_L$. This region, denoted by
$\mathcal{M}$, will be called the \emph{permitted region} (since
no $p$-positive distribution can exist outside this set). Some
examples of permitted regions, for different values of $d_c$, $K$
and $L$, all corresponding to $R=1/2$, are depicted in Fig.
\ref{fig:permitted-region}.
If $(\epsilon,\lambda_{L}) \in
\mathcal{M}$, then the polynomial $p\,(\cdot)$ could have in
principle 0, 2 or 4 real roots between 0 and 1 ($A=4$).

Let $\epsilon_{\min}^{\mathcal{M}}$ be the minimum value of
$\epsilon$ allowed for points within $\mathcal{M}$. If, for
$\epsilon=\epsilon_{\min}$, at least a $\lambda_L$ exists such
that $(\epsilon_{\min},\lambda_L)\in\mathcal{M}$, and such that
$p(\cdot)$ is not negative between 0 and 1, then it follows
$\epsilon_{\textrm{opt}}=\epsilon_{\min}$. On the contrary, if
$p(x)$ is negative for some $x$ between 0 and 1, for $\epsilon =
\epsilon_{\min}$ and for any admitted $\lambda_L$, it results
$\epsilon_{\textrm{opt}}>\epsilon_{\min}$. This is always the case
when $\epsilon_{\min}=0$. An approach to evaluate
$\epsilon_{\textrm{opt}}$ in this case is described next.

The proposed approach is based on this observation: If
$(\epsilon_{\textrm{opt}},\hat{\lambda}_L)$ is a solution of the
optimization problem (for some $\hat{\lambda}_L$), then at least
one $\overline{x} \in (0,1)$ must exist such that
\begin{align}
&p\,( \overline{x}; \epsilon_{\textrm{opt}}, \hat{\lambda}_L)=0 \label{eq:p0=0}\\
&p'(\overline{x}; \epsilon_{\textrm{opt}}, \hat{\lambda}_L)=0
\label{eq:p'0=0},
\end{align}
where the dependence of the coefficients of $p\,(\cdot)$ on
$\epsilon$ and $\lambda_{L}$ have been explicitly indicated. Then,
the search for the optimal distribution can be restricted to the
points $(\epsilon,\lambda_L) \in \mathcal{M}$ for which $p\,(
\overline{x}; \epsilon, \lambda_L)=0$ and $p'(\overline{x};
\epsilon, \lambda_L)=0$.

From these relationships, it is possible to obtain $\epsilon$ and
$\lambda_L$ as rational functions of $\overline{x}$, namely
$\epsilon=\epsilon(\overline{x})$ and
$\lambda_{L}=\lambda_{L}(\overline{x})$. Then, the technique
consists in plotting the trajectory of the point
$(\epsilon(\overline{x}),\lambda_{L}(\overline{x}))$ on the plane
$\epsilon - \lambda_L$, as well as the permitted region
$\mathcal{M}$. From this diagram it is usually possible to
univocally determine $\epsilon_{\textrm{opt}}$, as shown next with
an example.

\begin{figure}[t]
\begin{center}
\includegraphics[width=8.3 cm, angle=0]{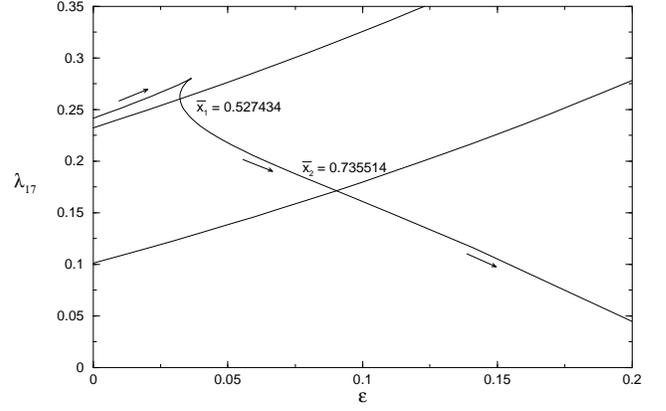}
\end{center}
\caption{Diagram showing the permitted region and the trajectory
of the point
$(\epsilon(\overline{x}),\lambda_{17}(\overline{x}))$, for
$R=1/2$, $d_c=7$, $K=5$ and $L=17$.} \label{fig:trajectory}
\end{figure}
\begin{figure}[t]
\begin{center}
\includegraphics[width=8.3 cm, angle=0]{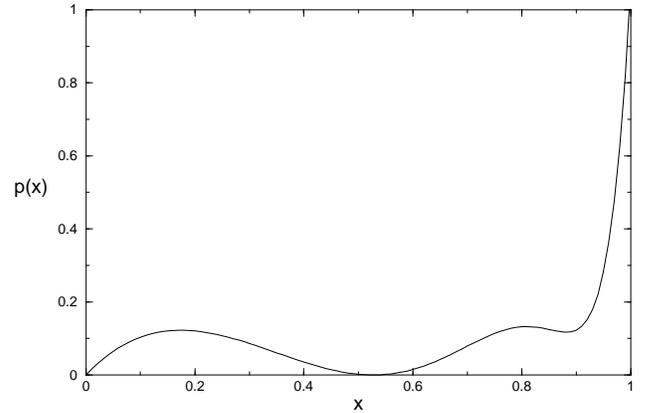}
\end{center}
\caption{Plot of $p\,(\cdot)$ for the optimal $R=1/2$ $p$-positive
distribution corresponding to $d_c=7$, $K=5$ and $L=17$.}
\label{fig:poly_plot}
\end{figure}
For instance, suppose $R=1/2$, $d_c=7$, $K=5$ and $L=17$. 
The trajectory of
$(\epsilon(\overline{x}),\lambda_{L}(\overline{x}))$ is depicted
in Fig. \ref{fig:trajectory}, together with a detail of the
permitted region. The trajectory of
$(\epsilon(\overline{x}),\lambda_{L}(\overline{x}))$ crosses the
permitted region for values of $\overline{x}$ between
$\overline{x}_1=0.527434$, and $\overline{x}_2=0.735514$. The
minimum-$\epsilon$ point on the trajectory segment
$\overline{x}_1\rightarrow\overline{x}_2$ is the intersection
point between the trajectory and the upper boundary of
$\mathcal{M}$. For this point, it results 
$\epsilon=\epsilon({\overline{x}_1})=0.032242$. The corresponding
polynomial $p\,(\cdot)$, whose graph is depicted in Fig.
\ref{fig:poly_plot}, does not assume negative values between 0 and
1. Hence, the minimum-$\epsilon$ point on the segment
${\overline{x}_1\rightarrow\overline{x}_2}$ corresponds to a
$p$-positive distribution: Then,
$\epsilon_{\textrm{opt}}=0.032242$. The threshold of the
corresponding degree distribution is
$\delta^*=[\lambda'(0)\rho'(1)]=(1-\epsilon_{\textrm{opt}})(1-R)=0.483879$,
the value of $\lambda_{17}$ can be obtained from the function
$\lambda_{17} = \lambda_{17}(\overline{x})$ 
and the values of $\lambda_3$ and $\lambda_5$ can be obtained from
the functions $\lambda_3 = \lambda_3(\epsilon,\lambda_{17})$ and
$\lambda_5 = \lambda_5(\epsilon,\lambda_{17})$.

\begin{table}[!t]
\caption{Optimal $R=1/2$ $p$-positive and not $p$-positive distributions for 
$d_c=\{6,7\}$ with $\lambda(x)=\lambda_2\,x +
\lambda_3\,x^2+\lambda_5\,x^{4}+\lambda_L\,x^{L-1}$.}\label{tab:distributions}
\begin{center}
\begin{tabular}{|c|c|c|c|c|}
\hline \multicolumn{1}{|c|}{$\,$} & \multicolumn{2}{|c|}{$d_c=6$, $L=10$} & \multicolumn{2}{|c|}{$d_c=7$, $L=15$}\\
\hline\hline $\,$ & $p$-pos. & DE & $p$-pos. & DE\\
\hline
$\lambda_2$ & 0.418913 & 0.415774 & 0.341501 & 0.339505\\
$\lambda_3$ & 0.167565 & 0.180916 & 0.142292 & 0.140214\\
$\lambda_5$ & 0.266696 & 0.248100 & 0.248395 & 0.259036\\
$\lambda_L$ & 0.146826 & 0.155210 & 0.267812 & 0.261244\\
\hline \hline
\multicolumn{5}{|c|}{\emph{Thresholds}}\\
\hline
$\delta^*$ & 0.477426 & 0.480325 & 0.488041 & 0.490947\\
\hline
\end{tabular}
\end{center}
\end{table}

By adopting a similar approach, we fixed $R=1/2$, $K=5$, and
looked for the best $p$-positive distribution for $d_c=6$ and
$d_c=7$. The optimal distributions were obtained for $L=10$ and
$L=15$, respectively. They are shown in Table
\ref{tab:distributions}, as well as the optimal check regular
distributions obtained by running the DE tool, under the same set
of constraints ($R$, $d_c$ and active variable degrees), but
without the $p$-positive constraint. The $p$-positive thresholds
are only slightly lower than the not $p$-positive counterparts,
and the degree distributions are quite similar. Again, the
$p$-positive distributions have the advantage represented by the
closed-form threshold expression $[\lambda'(0) \rho'(1)]^{-1}$.

\section{The Binomial Distributions are $p$-positive}\label{sec:binomial}
In \cite{shokrollahi99:new}, it was shown that capacity achieving
sequences of degree distributions for the BEC can be constructed
according to the binomial degree distribution. This is a check
regular distribution, whose variable distribution is in the form
$\lambda(x) =\sum_{i=2}^{L} \frac{\alpha \, {\alpha \choose i-1}
(-1)^{i}}{\alpha - L {\alpha \choose L} (-1)^{L+1} } x^{i-1}$,
where ${\alpha \choose N} = \alpha (\alpha-1) \dots
(\alpha-N+1)/(N!)$, and $\alpha = (d_c - 1)^{-1}$. Recognizing the
binomial degree distributions as part of the $p$-positive family,
the following theorem states that this family can achieve the BEC
capacity.

\medskip
\begin{theorem}\label{theo:binomial}
Any binomial degree distribution is $p$-positive.
\end{theorem}
\begin{proof}
Recall that, for $i=2,\dots,L-1$, the $(i-2)$-th coefficient of
$p\,(\cdot)$ is $p\,_{i-2}=\omega_i \lambda_i - (\omega_{i+1}+1)
\, \lambda_{i+1}$. For the binomial distribution, it is
$$
\omega_i \, \lambda_i = [(d_c-1)(i-1)-1]\,\frac{\alpha {\alpha
\choose i-1} (-1)^i}{\alpha - L {\alpha \choose L} (-1)^{L+1}}.
$$
Furthermore, it is
\begin{align*}
(\omega_{i+1} & +1)\, \lambda_{i+1} = (d_c-1)\,i\,\frac{\alpha
{\alpha \choose i} (-1)^{i+1}}{\alpha - L {\alpha \choose L}
(-1)^{L+1}} \\
\, & = (d_c-1) \, \frac{(i-1-\alpha)\,\alpha {\alpha \choose i-1 }
(-1)^i}{\alpha - L {\alpha \choose L} (-1)^{L+1}} \\
\, & = [ (d_c-1)\,(i-1) - \alpha\,(d_c-1) ]\,\frac{\alpha {\alpha
\choose i-1 } (-1)^i}{\alpha - L
{\alpha \choose L} (-1)^{L+1}} \\
\, & = [(d_c-1)(i-1)-1]\,\frac{\alpha {\alpha \choose i-1}
(-1)^i}{\alpha - L {\alpha \choose L} (-1)^{L+1}},
\end{align*}
where the second equality is due to the fact that ${\alpha \choose
i } = {\alpha \choose i-1} \frac{\alpha-i+1}{i}$, and the last
equality to the fact that $\alpha = (d_c-1)^{-1}$. Thus, for
$i=0,\dots,L-1$, $\omega_i \, \lambda_i = (\omega_{i+1} + 1) \,
\lambda_{i+1}$, i.e. $p\,_{i-2}=0$. It follows
$p\,(x)=\omega_L\,\lambda_L x^{L-2}$, which is always positive for
$x \in (0,1]$.
\end{proof}

\section{Conclusion}\label{section:conclusion}
In this paper, a special family of LDPC degree distributions has
been presented. The main feature of these distributions is the
possibility to express in closed form their decoding threshold on
the BEC, under iterative decoding. More specifically, the
threshold admits the simple closed form
$[\lambda'(0)\rho'(1)]^{-1}$. This family is a subset of the class
of check-regular distributions, and the distributions within this
family have been called $p$-positive distributions. A simple
necessary condition for a check regular distribution to belong to
this class has been obtained, by invoking some known results about
the real roots of polynomials.

Three additional properties of the proposed distributions have
been highlighted. The first one is their very good threshold under
some set of constraints. This property is not general, depending
on the specific imposed set of constraints. The second one is the
possibility, for particular structures of the variable
distribution, to optimize the distribution without necessarily
using the numerical optimization tools which are usually
exploited. The third one is the possibility to achieve the BEC
capacity with sequences belonging to the proposed family.

\section*{Acknowledgment}
This work was supported by the European Commission under project
FP6 IST-001812 "PHOENIX". The authors wish to thank Prof. Michele
Mulazzani, at the University of Bologna, for useful discussion
about polynomials and for suggesting the reference
\cite{barbeau03:polynomials}.
\end{document}